\documentclass[longauth]{aa} % for the long lists of affiliations 
\usepackage{graphicx}
\usepackage{natbib}
\usepackage{txfonts}
\begin{document}
   \title{A  search for debris disks in the \emph{Herschel} ATLAS\thanks{\emph{Herschel} is an ESA space observatory with science instruments provided by European-led Principal Investigator consortia and with important participation from NASA}}

   \subtitle{}

   \author{M.A. Thompson\inst{4} 
   \and D.J.B. Smith\inst{9}
   \and J.A. Stevens\inst{4}
   \and M.J. Jarvis\inst{4} 
   \and E. Vidal Perez\inst{3} 
   \and J. Marshall\inst{16}
   \and L. Dunne\inst{9}
   \and S. Eales\inst{2}
   \and G.J. White\inst{16,19} 
   \and  L. Leeuw\inst{15},  
   \and B. Sibthorpe\inst{13} 
   \and M. Baes\inst{3}
   \and E. Gonz{\'a}lez-Solares\inst{20}
 \and D. Scott\inst{22}
 \and
J. Vieiria\inst{21}
A. Amblard\inst{1}
\and
R. Auld\inst{2}
\and
D.G. Bonfield\inst{4}
\and
D. Burgarella\inst{5}
\and
S. Buttiglione\inst{6}
\and
A. Cava\inst{7,23}
\and
D.L. Clements\inst{8}
\and
A. Cooray\inst{1}
\and
A. Dariush\inst{2}
\and
G. de Zotti\inst{6}
\and
S. Dye\inst{2}
\and
S. Eales\inst{2}
\and
D. Frayer\inst{10}
\and
J. Fritz\inst{3}
\and
J. Gonzalez-Nuevo\inst{11}
\and
D. Herranz\inst{12}
\and
E. Ibar\inst{13}
\and
R.J. Ivison\inst{13}
\and
G. Lagache\inst{14}
\and
M. Lopez-Caniego\inst{12}
\and
S. Maddox\inst{9}
\and
M. Negrello\inst{16}
\and
E. Pascale\inst{2}
\and
M. Pohlen\inst{2}
\and
E. Rigby\inst{9}
\and
G. Rodighiero\inst{6}
\and
S. Samui\inst{11}
\and
S. Serjeant\inst{16}
\and
P. Temi\inst{15}
\and
I. Valtchanov\inst{17}
\and
A. Verma\inst{18}
}

\institute{
Dept. of Physics \& Astronomy, University of California, Irvine, CA 92697, USA
\and
School of Physics and Astronomy, Cardiff University,
  The Parade, Cardiff, CF24 3AA, UK
\and
Sterrenkundig Observatorium, Universiteit Gent, Krijgslaan 281 S9,
B-9000 Gent, Belgium
\and
Centre for Astrophysics Research, Science and Technology Research
Institute, University of Hertfordshire, Herts AL10 9AB, UK
\and
Laboratoire d'Astrophysique de Marseille, UMR6110 CNRS, 38 rue F.
Joliot-Curie, F-13388 Marseille France
\and
University of Padova, Department of Astronomy, Vicolo Osservatorio
3, I-35122 Padova, Italy
\and
Instituto de Astrof\'{i}sica de Canarias, C/V\'{i}a L\'{a}ctea s/n, E-38200 La
Laguna, Spain
\and
Astrophysics Group, Imperial College, Blackett Laboratory, Prince
Consort Road, London SW7 2AZ, UK
\and
School of Physics and Astronomy, University of Nottingham,
University Park, Nottingham NG7 2RD, UK
\and
National Radio Astronomy Observatory,  PO Box 2, Green Bank, WV  24944, USA
\and
Scuola Internazionale Superiore di Studi Avanzati, via Beirut 2-4,
34151 Triest, Italy
\and
Instituto de F\'isica de Cantabria (CSIC-UC), Santander, 39005, Spain
\and
UK Astronomy Technology Center, Royal Observatory Edinburgh, Edinburgh, EH9 3HJ, UK
\and
Institut d'Astrophysique Spatiale (IAS), B‰timent 121, F-91405 Orsay, France; and UniversitŽ Paris-Sud 11 and CNRS (UMR 8617), France
\and
Astrophysics Branch, NASA Ames Research Center, Mail Stop 245-6, Moffett Field, CA 94035, USA
\and
Department of Physics and Astronomy, The Open University, Milton Keynes, MK7 6AA, UK
\and
Herschel Science Centre, ESAC, ESA, PO Box 78, Villanueva de la
Ca\~nada, 28691 Madrid, Spain
\and
Oxford Astrophysics, Denys Wilkinson Building, University of Oxford, Keble Road, Oxford, OX1 3RH
\and
The Rutherford Appleton Laboratory, Chilton, Didcot OX11 0NL, UK
\and 
Institute of Astronomy, Madingley Road, Cambridge CB3 0HA, UK
\and
California Institute of Technology, 1200 East California Blvd., Pasadena, CA 91125, USA
\and
Department of Physics \& Astronomy, University of British Columbia, Vancouver, BC, V6T 1Z1, Canada
\and
Departamento de Astrof{\'\i}sica, Universidad de La Laguna, E-38205 La Laguna, Tenerife, Spain
}

   \authorrunning{M.A.~Thompson et al.}
   
   \titlerunning{A serendipitous search for debris disks in the \emph{Herschel} ATLAS}

   \date{}

% \abstract{}{}{}{}{} 
% 5 {} token are mandatory
 
  \abstract
   {}
   {We aim to demonstrate that the \emph{Herschel} ATLAS (H-ATLAS) is suitable for a blind and unbiased survey for debris disks by identifying candidate debris disks associated with main sequence stars in the initial science demonstration field of the survey.  We show that H-ATLAS  reveals a population of far-infrared/sub-mm sources that are associated with stars or star-like objects on the SDSS main-sequence locus. We validate our approach by comparing the properties of the most likely candidate disks to those of the known population.}
   {We use a photometric selection technique to identify main sequence stars in the SDSS DR7 catalogue and  a Bayesian Likelihood Ratio method  to identify H-ATLAS catalogue sources associated with these main sequence stars. Following this photometric selection we apply distance cuts to identify the most likely candidate debris disks and rule out the presence of contaminating galaxies using  UKIDSS LAS $K$-band images.  
  }
   {We identify 78 H-ATLAS sources associated with SDSS point sources on the main-sequence locus, of which two are the most likely debris disk candidates: H-ATLAS J090315.8 and H-ATLAS J090240.2. We show that they are plausible candidates by comparing their properties to the known population of debris disks. Our initial results indicate that bright debris disks are rare, with only 2 candidates identified in a search sample of 851 stars. We also show that H-ATLAS can derive useful upper limits for debris disks associated with Hipparcos stars in the field and outline the future prospects for our  debris disk search programme. }
   {}


   \maketitle
%
%________________________________________________________________

\section{Introduction}
The \emph{Herschel} ATLAS or H-ATLAS \citep{eales2010} is the largest Open Time Key Project on the \emph{Herschel} Space Observatory \citep{goran}, and will ultimately map over 500 square degrees with both the PACS and SPIRE instruments \citep{pog,griffin}. H-ATLAS is designed to  revolutionise our view of dust and  dust-obscured star formation by detecting $\sim$ 250\,000 galaxies in the far-infrared. The primary goal of the H-ATLAS is to study galaxy formation and evolution (see the other H-ATLAS articles in this volume), however the unrivalled sensitivity and wide-area coverage means that  H-ATLAS can also reveal dust in a range of more local (i.e.~within the Milky Way) environments. 
%The H-ATLAS fields lie at high galactic latitude to minimise cirrus contamination \citep{eales2010} and so any  Galactic dust that is detected must be no more than a few hundred parsecs distant.
At the sensitivity limits of H-ATLAS (a 5$\sigma$ threshold of 33 mJy at 250 $\mu$m measured from the science demonstration data, \citealt{pasquale2010,rigby2010}) this implies the potential to detect analogues of the well-known debris disks \citep[e.g.][]{holland1998,greaves1998} out to distances of between 20 and 150 pc.

 A search for debris disks in H-ATLAS  offers a powerful complement to those deeper and more targeted studies that are currently being undertaken with \emph{Herschel} (DUNES, DEBRIS and GASPS --- see publications in this volume), are set to be carried out with SCUBA-2 \citep[SUNS:][]{matthews2007} and have been perfomed by \emph{Spitzer} \citep[thoroughly described in][and references therein]{carpenter2009}. With its wide areal coverage H-ATLAS is a shallower tier to these  studies, but a tier with the potential to search a much larger number of stars for bright debris disks and with a large body of supporting high quality optical and infrared legacy data \citep{eales2010}. The H-ATLAS fields are covered by SDSS DR7  in $ugriz$ \citep{abazajian2009} and UKIDSS Large Area Survey  in $YJHK$ \citep{lawrence2007} with forthcoming deeper INT and VST KIDS optical data, VISTA VIKING in the near-infrared, WISE in the mid-infrared and GMRT radio continuum. The supporting optical \& infrared data allows straightforward selection of main sequence stars in the H-ATLAS fields via the main sequence colour locus \citep{covey2007} and the use of techniques to exclude contaminating background galaxies, such as  the inspection of deep $K$-band images for extended objects  and the use of the FIR-radio correlation \citep{carilli1999}. Finally, a debris disk search in H-ATLAS is completely serendipitous and carried out in parallel with the primary science programme.

In Sect.\ref{sect:technique} we show that the full H-ATLAS survey will allow us to search $\sim$ 10\,000 main sequence stars for the presence of bright debris disks analogous to Beta Pictoris and $\sim$ 1\,000 stars for  Fomalhaut analogues. This large search sample means that H-ATLAS will be much more sensitive to rarities in the debris disk population than targeted surveys, leading to stringent tests of stochastic disk evolution models \citep{wyatt2008} and potentially uncovering bright and/or cold disks that may have undergone recent disruptive events \citep[e.g.][]{lisse2007}. H-ATLAS will allow us to answer questions such as how frequent are bright systems such as Beta Pictoris or HR 4796 and is there an upper limit to the amount of debris formed during disk evolution? 
%The more distant disks ($> 30$ pc) discovered by  H-ATLAS will have radii well-matched to interferometric follow-up by ALMA.  

%As the H-ATLAS survey area is also covered by SDSS, there is also the possibility of undertaking surveys towards SDSS spectroscopically selected samples such as White Dwarfs \citep{eisenstein2006} or Ultracool Dwarfs \citep{west2008}. 

%In this Letter we describe the first results of our search programme, outlining  our photometric selection techniques in Sect.~\ref{sect:technique} and  presenting our highest significance disk candidates in Sect.~\ref{sect:results}. We discuss the implications of our results and outline the future H-ATLAS debris disk search programme in Sect.~\ref{sect:conclusions} 

%__________________________________________________________________

\section{Photometric selection of main sequence stars in the H-ATLAS Science Demonstration Field}
\label{sect:technique}

The H-ATLAS science demonstration (SD) field occupies roughly 16 square degrees and is centred at 09:05:30.0 $+$00:30:00 (J2000). Descriptions of the PACS and SPIRE images obtained in parallel mode and the data reduction procedure used can be found in \citet{ibar2010} and \citet{pasquale2010} respectively. Point sources were identified within the images using a combination of PSF filtering, Gaussian fitting and aperture photometry \citep{rigby2010}. The median 5$\sigma$ noise values (including confusion noise) in the SPIRE 250, 350 \& 500\,$\mu$m maps are 33, 36 \& 45\,mJy/beam respectively. The PACS images are roughly a factor 2 noiser than predicted with 5$\sigma$ noise levels of 105 and 138 mJy/beam at 100 and160\,$\mu$m respectively. The H-ATLAS source catalogue for the SD field \citep{rigby2010} contains 6878 band-merged sources with flux density $>5\sigma$ in at least one of the 5 H-ATLAS bands (100, 160, 250, 350, 500 $\mu$m). The catalogue excludes higher noise regions at the edge of the map and thus covers an effective area of 14.5 square degrees. Each source in the H-ATLAS catalogue has been matched to the corresponding most likely SDSS DR7 object  within a 10\arcsec\ search radius. This matching process is described further in \citet{smith2010} and is an implementation of the  Bayesian Likelihood Ratio technique of \citet{sutherland1992}.

In order to search for debris disks in this catalogue, we must first identify a sample of main sequence stars. We use the main sequence colour locus identified by \citet{covey2007},  which constrains the location of main sequence stars in SDSS colour space. This approach allows us to maximise our search sample by going to faint magnitudes and  takes advantage of the well-calibrated and well-understood SDSS optical photometry. We use a 4-dimensional main sequence locus as described in \citet{kimball2009} rather than the full 7-dimensional SDSS+2MASS locus of \citet{covey2007}. The infrared excess of a warm debris disk at $K$-band can move our target stars away from the nominal locus, and as our aim is to identify stars that are potential debris disk hosts, we thus do not use 2MASS or UKIDSS colours in our photometric selection.

\begin{figure}
\includegraphics*[angle=-90,width=9 cm]{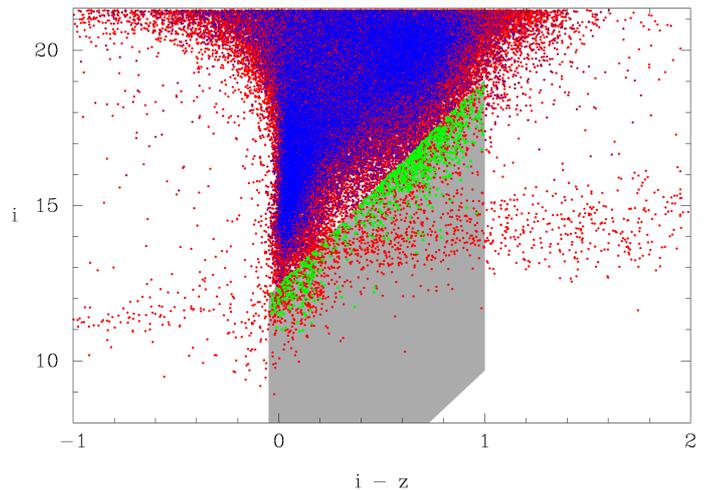}
\caption{SDSS $i$ vs.~$i-z$ colour magnitude diagram of SDSS point sources (i.e.~with probPSF = 1) in the H-ATLAS SD field. A magnitude cut of $i<21.3$ has been applied to exclude sources with large photometric error. Dots indicate field SDSS point sources (red) and point sources within the main sequence stellar locus (blue). The grey shaded polygon indicates the colour-magnitude region occupied by main sequence stars between  distances  of 4 and 200 pc (calculated using the \citet{davenport2006} photometric parallax relation). Stars within the main sequence locus that fall into this box are shown as larger green points for clarity.}
\label{fig:sdss_cmd}
\end{figure}

Figure~\ref{fig:sdss_cmd} shows a colour-magnitude diagram of the 180\,000 star like objects (selected with 'probPSF=1') detected by  SDSS DR7  in the H-ATLAS SD field. The general population  are shown by red dots and those falling within 2 `units' of the 4-dimensional main sequence color locus defined by \citet{kimball2009} are shown in blue. Note that the colour locus is 4-dimensional and Fig.~\ref{fig:sdss_cmd} shows only a 1D cut through the locus. Figure~\ref{fig:sdss_cmd} shows that the bulk of the main sequence stellar population detected by SDSS is comprised of faint and likely distant halo stars. However, there are a substantial number of relatively bright ($11 < i <17$) and hence likely nearby stars for which  it is possible that H-ATLAS could detect associated debris disks. 

We estimate the maximum distance to which H-ATLAS could be sensitive to debris disks by scaling from the  \emph{Spitzer} MIPS and SCUBA SEDs of known examples (Beta Pictoris, \citealt{rebull2008}; Epsilon Eridani, \citealt{backman2009}; HR 8799, \citealt{lisse2007}; Fomalhaut, \citealt{stapelfeldt2004}, Vega \& HR 4796, \citealt{sheret2004}). The most sensitive wavelength of H-ATLAS is 250 $\mu$m (at which the stellar photospheric contribution is minimal) and we find that the maximum distances for these debris disk analogues are: HR 4796 $<$200 pc, Beta Pictoris $<$150 pc, Fomalhaut $<$80 pc, Vega/HR 8799 $<$50 pc and Epsilon Eridani $<$20 pc. We indicate the colour-magnitude region in which  main sequence dwarf stars at a distance of 4--200 pc should lie in Fig.~\ref{fig:sdss_cmd} by a grey shaded box, calculated  using the distance modulus and  the $M_{i}$ vs.~$(i-z)$ photometric parallax relation for dwarf stars from \citet{davenport2006}. Note that late M and L dwarfs follow a shallower relation for $(i - z) > 1.26$ \citep{west2005}, and also that we have extrapolated the \citet{davenport2006} relation to $(i  - z) < 0.2$ to account for the bluer stars in the sample.

In the H-ATLAS SD field  for maximum distances of 200, 150, 80 and 50 pc (i.e.~sensitive to HR 4796, Beta Pictoris, Fomalhaut and Vega analogues) we find a total of 851, 340, 31 and 9 stars respectively on the main sequence locus. There are no stars on this locus nearer than 20 pc in the SD field, which is likely an effect of the SDSS  becoming saturated for near, bright stars. Such stars can be obtained from the Tycho-2 \& Hipparcos catalogues \citep{hog2000, vanleeuwen2007}, although in this paper we focus upon the  larger and better selected SDSS sample of main sequence stars. Assuming similar stellar densities in the remaining 550 square degrees  that will be mapped in the full survey, H-ATLAS will thus encompass a search sample on the order of 10\,000 main sequence stars for the brightest debris disks (HR 4796 and Beta Pictoris analogues) and on the order of  300--1000 stars for Fomalhaut and Vega analogues.

\section{Candidate debris disks in the H-ATLAS SD Field}
\label{sect:results}

\begin{figure}
\includegraphics[angle=-90,width=9cm]{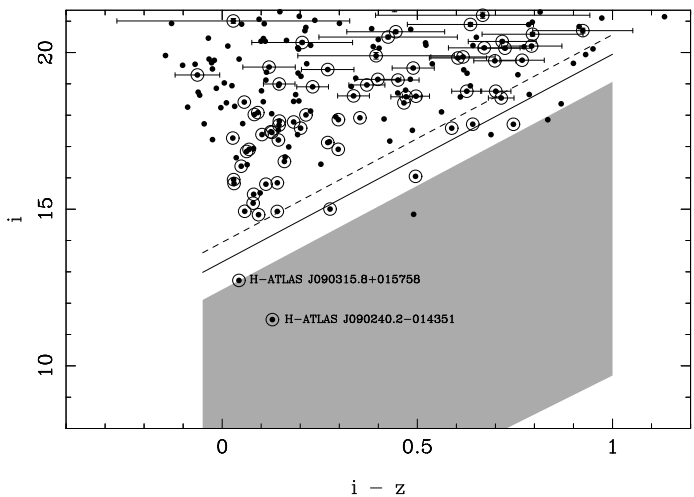}
\caption{SDSS $i$ vs.~$i-z$ colour magnitude diagram of SDSS point sources with high reliability matches to H-ATLAS catalogue sources. Objects whose SDSS colours place them on the stellar locus are identified by open circles. Thre grey shaded region indicates the colour-magnitude region occupied by stars between 4 and 200 pc as in Fig.~\ref{fig:sdss_cmd}. Solid and dashed lines indicate distances of 300 \& 400 pc respectively. The two candidate debris disks are identified by their H-ATLAS catalogue ID.}
\label{fig:atlas_cmd}
\end{figure}

We identify candidate debris disks by taking a sample of H-ATLAS sources from the catalogue that have a high reliability match ($>$80\% reliability) to SDSS DR7 catalogued sources \citep{smith2010}. There are 2334 sources that pass this criterion. We then filter this list to only include SDSS point sources ('probPSF=1') and identify point sources on the 4-dimensional main sequence colour locus described in the previous section.  We find a total of 204 H-ATLAS sources matched to SDSS sources with 'probPSF=1', of which 78 sources fall within the main sequence locus (see Fig.~\ref{fig:atlas_cmd}). Within this sample we expect considerable contamination from dust obscured QSOs or unresolved galaxies whose optical colours are reddened into the main sequence locus \citep{ivezic2002}. Indeed, the sample have a median  $r$-band magnitude of 19.8, fainter than the value at which galaxies dominate over stars \citep{covey2007}. Unsurprisingly due to their optical colours, none of the H-ATLAS main sequence locus objects have measured SDSS spectroscopic redshifts which would allow us to select out QSOs and unresolved galaxies. The inclusion of UKIDSS near-IR colours in the selection would aid in this discrimination \citep{ivezic2002}, but as we mention in the previous section, would also select against possible $K$-band excess in debris disk stars.

To identify candidate disks in our sample we apply a photometric distance cut to select  the brightest and nearest objects that are least likely to be QSOs or unresolved galaxies and the most likely to be debris disks.  We apply an initial photometric distance cut of 200 pc to select the most likely candidate disks, though Fig.~\ref{fig:atlas_cmd} shows that there are a further 7 objects at 200--400 pc distance that could be  massive or luminous disk candidates. The candidates that pass the main sequence colour locus and  photometric distance tests are then  finally subject to a detailed inspection of SDSS DR7 $ugriz$ and UKIDSS LAS $YJHK$ images to reject the presence of possible contaminating galaxies, and the wider field of the H-ATLAS SPIRE images to search for contaminating cirrus. We stress that our search technique reveals \emph{candidate} disks. Spectroscopic confirmation of the host star spectral types, more accurate spectrophotometric distances,  higher resolution PACS or SCUBA-2/ALMA imaging, or scattered light imaging are required to confirm these objects as debris disks and to better constrain their physical properties such as temperature and mass.

\begin{figure}\centering
\includegraphics*[width=7cm,trim=50 35 100 40]{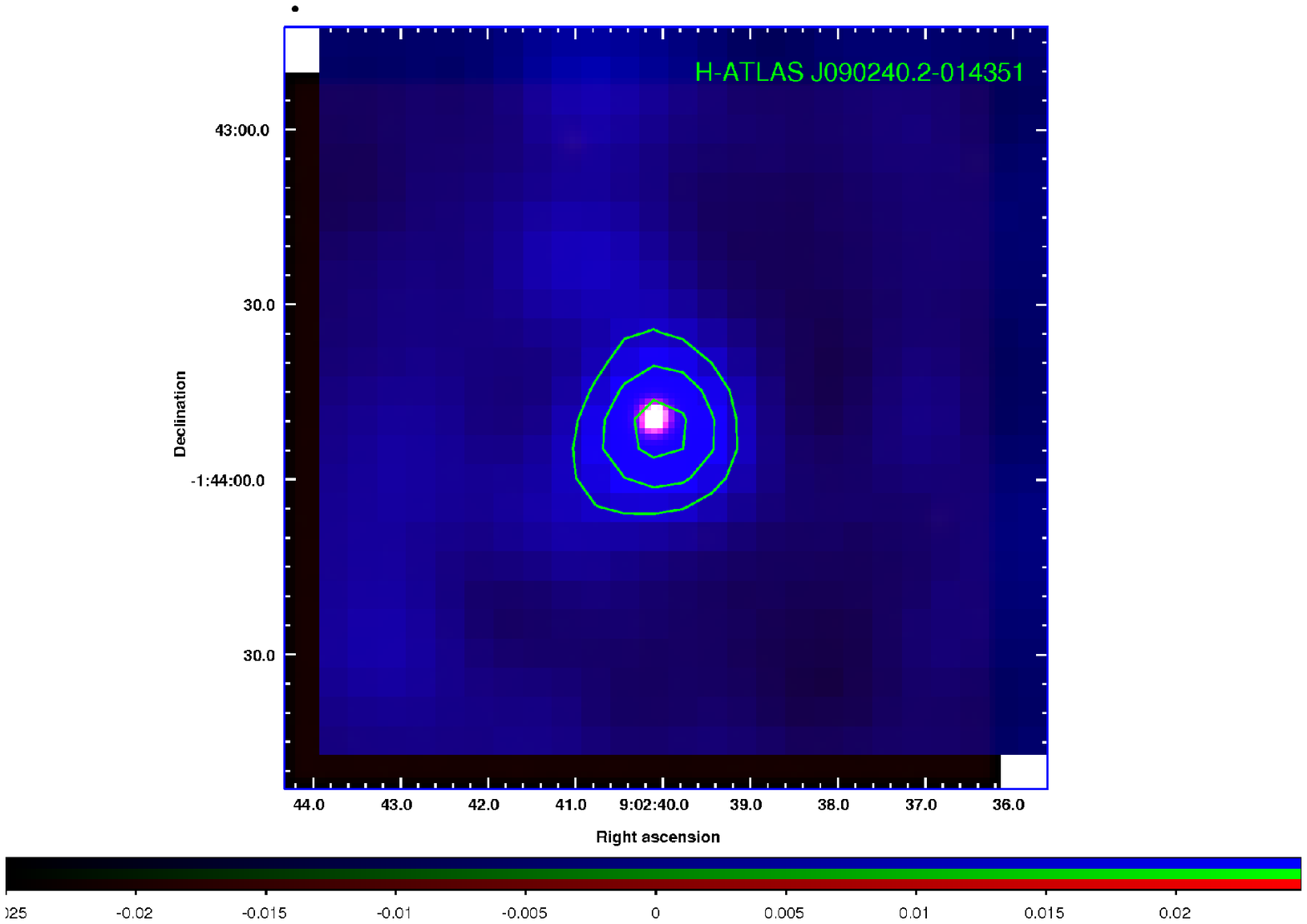}
\vspace*{0.1cm}\includegraphics*[width=7cm,trim=50 35 100 40]{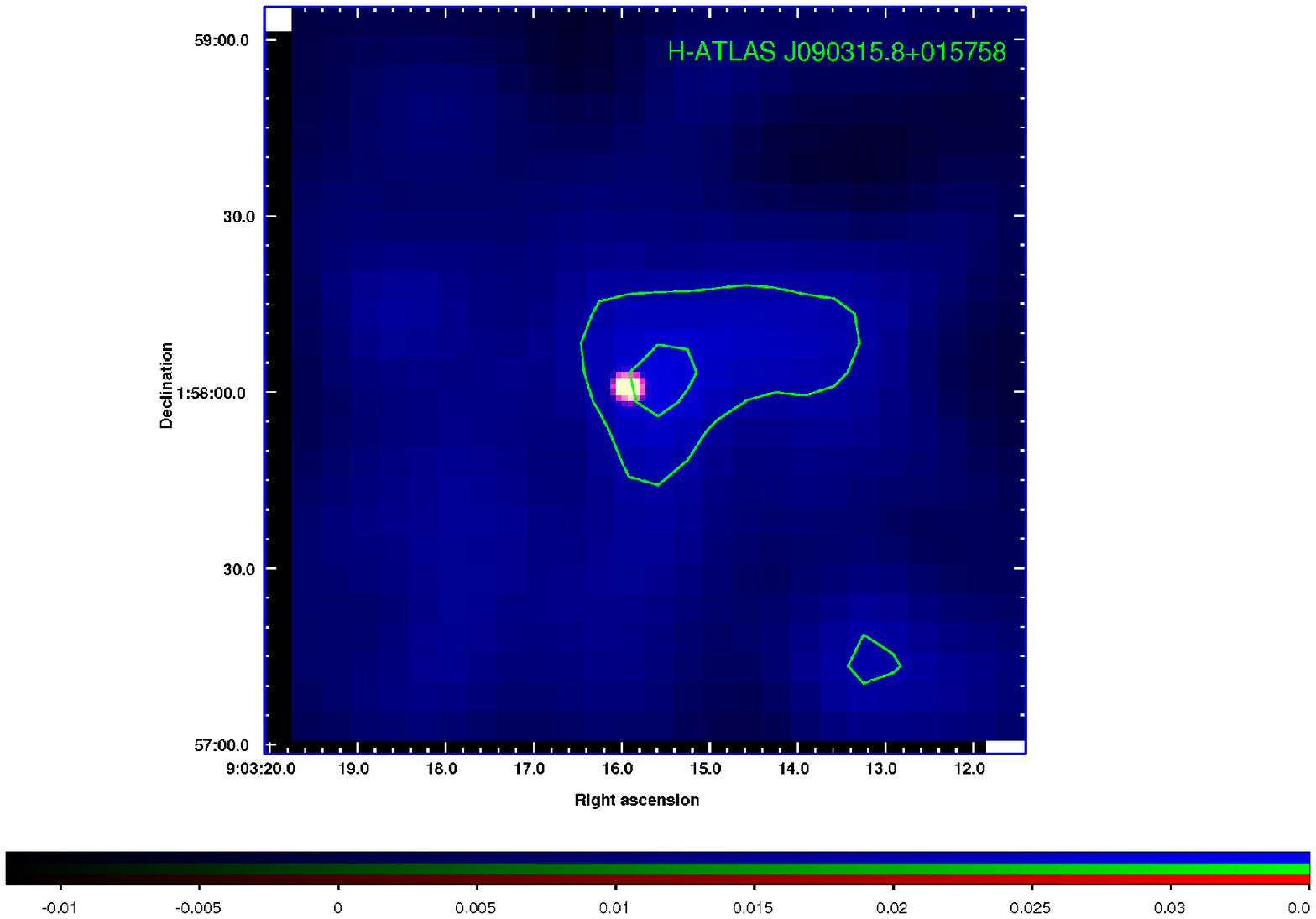}
\caption{Three colour images of the two candidate debris disks detected by H-ATLAS. Red \& green channels are 2MASS J \& K$_{\rm s}$ respectively. The blue channel is SPIRE 250$\mu$m, smoothed by a 3 pixel Gaussian kernel to increase the signal to noise ratio. Contours are of 250 $\mu$m emission starting at 3$\sigma$ and spaced by 2$\sigma$.}
\label{fig:disks}
\end{figure}

We focus our following discussion on the two closest candidate disks found within a  photometric distance cut of 200 pc: H-ATLAS J090315.8+015758 and H-ATLAS J090240.2$-$014351.  As we will show, the physical properties of these objects are within the spectrum of known debris disks and the H-ATLAS detections are thus consistent with a debris disk hypothesis. We summarise the FIR and stellar properties of  H-ATLAS J090315.8 and H-ATLAS J090240.2 in Table \ref{tbl:disks} and present three colour  images of the disks in Fig.~\ref{fig:disks}. The $g-i$ colours of the host stars imply spectral types of K2 and G5 respectively \citep{covey2007}. Their 2MASS colours or brightnesses are inconsistent with those of giant stars ($J-H$ \& $H-K_{s}$  for both stars is $<$0.3 and $\le0.1$ respectively). The photospheric flux of these stars at 250 $\mu$m is of the order of a few $\mu$Jy and so we can be confident that the 250 $\mu$m emission is a genuine excess over the stellar spectrum.  Both disks are unresolved at 250 $\mu$m, although  H-ATLAS J090315.8  shows a marginal extension  to the West. H-ATLAS J090240.2 on the other hand is compact and centred on the star's position to within the pointing accuracy of \emph{Herschel}. Inspection of  UKIDSS $K$-band images shows that background galaxy contamination is unlikely.

Both candidate disks are detected in the 250 \& 350 $\mu$m SPIRE bands, but not at 100, 160 or 500 $\mu$m (both disks have emission significant at the 2$\sigma$ level at 500 $\mu$m). With only two flux points it is difficult to constrain either the fractional luminosity or the temperature of the disk. We estimate temperature by fitting a fixed $\beta$=0.5 modified blackbody \citep[e.g.][]{wyatt2005} to the 110, 160 $\mu$m upper limits and 250, 350 \& 500 $\mu$m flux measurements. As the 110 and 160 $\mu$m flux are only an upper limit the derived temperatures should be considered as strict upper limits. We note that as our photometric distance estimates are only good to within $\sim$50\% \citep{davenport2006}, the error in derived disk mass is largely dominated by distance rather than temperature. Using the standard techniques outlined in \citet{holland1998} and \citet{sheret2004} we derive masses for the two candidate disks of 0.06--0.13 M$_{\oplus}$ and 0.5--1.3M$_{\oplus}$ for  H-ATLAS J090315.8 and H-ATLAS J090240.2 respectively. The mass of both candidate disks is within the observed range of known disks \citep{wyatt2008,lisse2007,sheret2004}, comparing specifically to HD 12167 with a mass of 1 M$_{\oplus}$ and $\beta$ Pic with a mass of 0.1 M$_{\oplus}$. \citep{sheret2004}

\begin{table*}
\caption{FIR and stellar properties of the candidate disks \& their associated stars. Spectral types are estimated from the $(g - i)$ colour relation of \citet{covey2007} \& distances from the photometric parallax from \citet{davenport2006}. Temperatures are estimated by a greybody fit to the 250, 350 \& 500 $\mu$m fluxes and the upper limits of 140 mJy at 160 $\mu$m, and 105 mJy at 110 $\mu$m with a fixed $\beta$= 0.5 \citep{wyatt2005}. Masses are calculated using the standard technique with a mass absorption coefficient  $\kappa$ = 1.7 cm$^{2}$\,g$^{-1}$ at 850 $\mu$m \citep{sheret2004}.}
\label{tbl:disks}
\begin{tabular}{llccccccc}\hline
H-ATLAS ID & SDSS ID &  Sp.~type & Distance (pc) &  $F_{250}$ (mJy) & $F_{350}$ (mJy) & F$_{500}$ (mJy) & T (K) & Dust mass ($M_{\oplus}$)\\\hline
 J090315.8+015758 & J090315.91+015800.9 &  K2 & 90--130 & $51\pm14$ & $53\pm14$ & $19\pm9$ & 65 & 0.06--0.13 \\
 J090240.2$-$014351 & J090240.10-014349.7 & G5 & 190--290 & $86\pm12$ & $48\pm7$ & $17\pm9$ &  60 & 0.5--1.3 \\
\end{tabular}
\end{table*}

%\begin{figure}
%\centering
%\includegraphics*[width=7cm,trim=350 20 650 40]{hatlas403_ukidssK_PSW.eps}
%\caption{UKIDSS LAS K band images of . Black contours are of 250 $\mu$m emission from ATLAS, starting at 4$\sigma$ and increasing by 4$\sigma$ (25 mJy). The white contour is a negative 4$\sigma$ contour to %give a guide to the significance of the lowest contour plotted. Note the presence of faint extended objects in the K band images to the SW of TYC 226-2344-1.}
%and the SE of TYC 4878-402-1.}
%\label{fig:kdisks}
%\end{figure}

\section{Conclusions \& future prospects for H-ATLAS debris disk searches}
\label{sect:conclusions}

 We have described a search method for debris disks in the \emph{Herschel} ATLAS and present the  two most likely candidate disks in the H-ATLAS SD field: H-ATLAS J090315.8 and H-ATLAS J090240.2, $\sim$ 0.1and 1 M$_{\oplus}$ mass candidate disks around K0 and G5 stars respectively. We also identify a further population of 76 SDSS point sources that are associated with FIR/sub-mm emission and whose optical colours place them on the main sequence locus. The majority of these objects are likely to be dust-obscured QSOs or unresolved galaxies \citep[e.g.][]{ivezic2002}, although the  brighter and nearer (7 objects lie within a photometric distance of 400 pc) may be potentially luminous debris disks. Follow-up spectroscopy is required to investigate these hypotheses. What is clear from the H-ATLAS SD field is that bright disks are rare --- we have searched 851 stars within 200 pc for FIR/sub-mm emission and find only two candidate disks brighter than 33 mJy at 250 $\mu$m. Our search finds a much lower fraction of candidate debris disks than previous \emph{Spitzer} and SCUBA studies \citep[e.g.][]{carpenter2009,hillenbrand2008,wyatt2008}, which typically find a 7--14\% disk detection fraction. However we note that direct comparisons between these detection fractions should not be made as H-ATLAS is flux-limited rather than volume-limited and we do not reach the exquisite photospheric signal-to-noise levels of the \emph{Spitzer} studies \citep{carpenter2009}. Hence H-ATLAS is likely to only detect the bright end of the debris disk population. 

The search that we present in this Letter is a forerunner to a much more extensive programme that will be carried out in the future. The full H-ATLAS will contain a search sample of $\sim$10\,000 photometrically selected main sequence stars out to 200 pc, allowing us to place  more stringent limits upon the frequency of the bright end of the debris disk population. With a large and well-selected sample of main sequence stars covering a range of spectral types we will also be able to carry out a stacking analysis \citep[e.g.][]{kurczynski2010} to determine the statistical frequency of disk occurrence as a function of spectral type. We also plan to include bright stars from the Tycho-2 \& Hipparcos catalogues, which will extend our search to nearer main sequence stars. A preliminary search of the Tycho-2  catalogue indicates that none of the 569 stars found within the H-ATLAS SD field are associated with detectable debris disks. For the 7 nearest Hipparcos main sequence stars within our field (which lie between 30 \& 80 pc) this means that H-ATLAS can place upper limits on their disk masses of 0.01--0.07 M$_{\oplus}$ (or between 0.8 and 5.7 Lunar masses of dust, assuming $T_{\rm dust}$ = 40 K). Finally,  H-ATLAS   has significant legacy potential for the GAIA mission \citep{lindegren2008}, which will determine distances and spectral types for all stars brighter than $r=20$ in the H-ATLAS fields. In combination with GAIA parallaxes, H-ATLAS will be able to determine precise disk mass upper limits for a large sample of stars.

\begin{acknowledgements}
Funding for the SDSS and SDSS-II has been provided by the Alfred P. Sloan Foundation, the Participating Institutions, the National Science Foundation, the U.S. Department of Energy, the National Aeronautics and Space Administration, the Japanese Monbukagakusho, the Max Planck Society, and the Higher Education Funding Council for England. 
The UKIDSS project is defined in \citet{lawrence2007}. UKIDSS uses the UKIRT Wide Field Camera \citep[WFCAM; ][]{casali2007}. The photometric system is described in \citet{hambly2008}, and the calibration is described in \citet{hodgkin2009}. The pipeline processing and science archive are described in \citet{hambly2008}. MAT would like to thank two of our undergraduate project students, Sam Richards \& Max Podger, who carried out initial database searches and also David Pinfield and Ralf Napiwotski for  discussions on low mass stars.
\end{acknowledgements}

\end{document}